\documentclass[aps,pra,reprint,showpacs,superscriptaddress,twocolumn]{revtex4-2}

\usepackage{amsfonts}
\usepackage{amsmath}
\usepackage{amssymb}
\usepackage[T1]{fontenc}
\usepackage{hyperref}
\usepackage[utf8]{inputenc}
\usepackage{orcidlink}
\usepackage{physics}
\usepackage{times}
\usepackage{xcolor}

\hypersetup{
 colorlinks=true,linkcolor=blue,citecolor=blue,
 filecolor=blue,urlcolor=blue,breaklinks=true
}

\RequirePackage{color}

\newcommand{\air}{
 Air Institute,
 Carr. de la Aldehuela,
 Zamora,
 49022 Castilla y León, Spain}

\newcommand{\qpqi}{
  QPQI Group,
  Universidade Estadual de Ponta Grossa,
  84030-900 Ponta Grossa, Paraná, Brazil
}

\newcommand{\uepg}{
 Departamento de Matemática e Estatística,
 Universidade Estadual de Ponta Grossa,
 84030-900 Ponta Grossa, Paraná, Brazil
}

\newcommand{\ufma}{
Programa de P\'os-Gradua\c c\~ao em F\'{\i}sica \& Coordena\c c\~ao do Curso de F\'{\i}sica -- Bacharelado, Universidade Federal do Maranh\~{a}o, 65085-580 S\~{a}o Lu\'{\i}s, Maranh\~{a}o, Brazil}

\newcommand{\orcidvinicius}{\orcidlink{0000-0002-1768-8783}}
\newcommand{\orcidedilberto}{\orcidlink{0000-0003-3552-8780}}
\newcommand{\orcidfabiano}{\orcidlink{0000-0002-8882-2169}}

\begin{document}

\title{
  Universal critical $g$ factor for spin-1/2 Aharonov-Bohm bound states%
}

\author{Vinícius Salem\orcidvinicius}
\email{vsalem@air-institute.com}
\affiliation{\air}

\author{Edilberto O. Silva\orcidedilberto}
\email{edilberto.silva@ufma.br}
\affiliation{\ufma}

\author{Fabiano M. Andrade\orcidfabiano}
\email{fmandrade@uepg.br}
\affiliation{\qpqi}
\affiliation{\uepg}

\date{\today}

\begin{abstract}
Whether the spin-1/2 Aharonov-Bohm Hamiltonian supports bound states has remained controversial since Hagen concluded that no such states exist for a Dirac particle. Here, we show that Hagen's solution corresponds to a particular member of the one-parameter family of self-adjoint extensions that describe the singular Zeeman interaction, and we identify which member the microscopic physics selects. Matching the extension parameter to a finite-radius flux tube yields $\nu=0$  precisely at $g=2$, and it does so in every flux sector. The critical $g$ factor is therefore universal, $g_c=2$, and depends on neither the flux sector nor the regularization radius. The flux sector controls the depth of the bound state energy, and we obtain a closed-form expression showing that binding deepens markedly with the integer part of the flux. Hagen's conclusion is thus confirmed and sharpened: it is not a statement about one arbitrary member of a family of extensions, but about the member that the physics selects when the magnetic moment takes its Dirac value, and an anomalous moment is a necessary and sufficient condition for binding in the singular channel.
\end{abstract}

\maketitle

\section{Introduction}

The Aharonov-Bohm (AB) effect is one of the most fundamental
manifestations of the physical significance of electromagnetic
potentials in quantum mechanics
\cite{ehrenberg1949refractive,PR.1959.115.485,tonomura1986evidence,bezerra1989gravitational,Tonomura1992,PhysRevA.86.040101}.
For spin-$1/2$ particles, however, the Zeeman coupling between the
magnetic moment and the magnetic flux introduces a singular point
interaction, whose proper treatment remains a fundamental and
controversial problem.
Whether this singular interaction supports bound states has been the
subject of continuous debate for more than three decades
\cite{CMP.124.229.1989,PRD.1989.40.1346,JMP.1995.36.5453,Book.1995.Jackiw,
Moroz1996,PRL.1990.64.503,JPA.1993.26.7637,PRD.2012.85.041701,
AoP.339.510.2013,JMP.1998.39.47,LMP.1998.43.43,Coutinho1992,Pankrashkin2011}.
While analyses based on self-adjoint extensions suggested that bound
states may occur under suitable boundary conditions
\cite{CMP.124.229.1989,PRD.1989.40.1346,JMP.1995.36.5453,Book.1995.Jackiw,Moroz1996},
Hagen argued that for the Dirac value of the magnetic moment (or $g$
factor) of $g=2$, every physically acceptable regularization excludes
bound states  \cite{PRL.1990.64.503}.
A central unresolved issue is whether the onset of bound state
formation is governed by a universal physical criterion or by the
microscopic regularization adopted for the singular interaction.
Subsequent investigations employing finite-radius flux tube
regularization established the existence of bound states for particles
with anomalous magnetic moments ($g>2$).
However, the relationship between these results and the corresponding
self-adjoint extension has remained unclear
\cite{Moroz1996,JPA.1993.26.7637,PRD.2012.85.041701,AoP.339.510.2013}.
In particular, the number of bound states was found to depend on the
adopted regularization \cite{Moroz1996}, and no closed-form criterion
was available to identify, channel by channel, the threshold
separating the two regimes.

Singular point interactions, common in systems with topological
defects \cite{Book.2000.Vilenkin,AoP.1992.216.1}, require a
mathematical treatment that does not impose artificial restrictions.
The self-adjoint extension formalism \cite{Book.1975.Reed.II}
classifies all the admissible quantum dynamics associated with point
interactions
\cite{NPB.1989.328.140,Book.2004.Albeverio,JMP.1985.26.2520}, but it
does not determine which member of this family is physically realized.
Consequently, the existence or absence of bound states cannot be
inferred solely from mathematical consistency, as it requires a
physical mechanism that selects the appropriate extension.
The microscopic structure of the interaction must ultimately determine
that choice. We show that the finite-radius regularization considered
here, combined with the magnetic moment through the $g$ factor,
provides this selection mechanism, thus uniquely determining the
self-adjoint extension that describes the short-distance dynamics.

In this work, we carry out that selection explicitly.
By modeling the flux tube as a solenoid of finite radius $R$ and
matching the interior and exterior solutions at $r=R$, we obtain the
self-adjoint extension parameter $\nu$ in closed form.
Once it is retained, the matching yields $\nu=0$ for the extension
underlying Hagen's scattering solution \cite{PRL.1990.64.503}, which
is exactly associated with $g=2$, and it does so in every flux sector.
The critical $g$ factor is therefore universal, $g_c=2$, and is
independent of both the flux sector and the regularization radius.
An anomalous magnetic moment is thus a necessary condition for the
formation of a bound state in the singular channel.
We also obtain a closed-form expression for the binding energy, which
shows a pronounced enhancement with $|N|$, the modulus of the integer
part of the flux.

This work is organized as follows.
In Section \ref{sec:abh}, we present the Hamiltonian of the system and
its formal solutions.
In Section \ref{sec:sae}, we summarize the self-adjoint extension
approach as used in this work, discussing the family of boundary
conditions.
In Section \ref{sec:ss}, we solve the scattering problem, showing how
our general result reduces to Hagen's original result.
In Section \ref{sec:bs}, we perform the physical selection of the
extension and derive the binding threshold and the flux-sector
dependence of the binding energy.
In Section \ref{sec:exp}, we provide some experimental implications.
Finally, in Section \ref{sec:conc}, we present our conclusions.

\section{The spin-1/2 AB Hamiltonian}
\label{sec:abh}
The Dirac equation for a charged particle of mass $M$, possessing an
anomalous magnetic moment, with the inclusion of external
electromagnetic fields, is written as ($\hbar=c=1$)
\begin{equation}
  \label{dp}
  \left[
    i\gamma ^{\mu }
    \left(\partial _{\mu }+ieA_{\mu }\right)
    -a \frac{\mu_{B}}{2}\sigma ^{\mu \nu }F_{\mu \nu }-M
  \right] \Psi =0,
\end{equation}
with $A_{\mu}=(A_{0},-\mathbf{A})$ representing the four-vector
potential of the associated electromagnetic field, $a = (g-2)/2$ is
the $g$ factor anomaly, $F_{\mu \nu }$ is the electromagnetic field
tensor, whose components are $F_{0j}=E^{j}$, $F_{mn}=\epsilon
_{mnp}B^{p}$, and $\mu _{B}=e/2M$ is the Bohr magneton.
The operator $\sigma ^{\mu \nu }$ is given in terms of Dirac matrices
as
$\sigma^{\mu \nu }
=i(\gamma ^{\mu}\gamma ^{\nu }-\gamma ^{\nu }\gamma^{\mu })/2
=i[\gamma ^{\mu },\gamma ^{\nu }]/2,$ with
$\sigma ^{0j}=i\alpha ^{j},$ $\sigma ^{ij}=-\epsilon _{ijk}\Sigma ^{k}$
being its components.

The nonrelativistic reduction of the Dirac equation follows from
the decoupling of the (3+1)-dimensional Dirac equation for
the specialized case where $\partial _{3}=z=0$ and $A_{3}=0$, together
with the representation of the Dirac matrices in terms of the Pauli
matrices, $\gamma ^{0}=\sigma ^{3}$, $\gamma ^{1}=i\sigma ^{2}$,
$\gamma ^{2}=-is\sigma ^{1}$, where the parameter $s$, which is twice
the spin value, was introduced to characterize the two spin states:
$s=+1$ for spin ``up'' and $s=-1$ for spin ``down''.
As a result of this procedure, we obtain two uncoupled two-component
equations, such as those implemented in
Refs. \cite{PRD.1978.18.2932,NPB.1988.307.909,PRL.1989.62.1071}.
Thus, Eq. \eqref{dp}, after taking the nonrelativistic limit, reads
\cite{Book.2012.Itzykson}
\begin{equation}
  \left(\mathbf{p}-e\mathbf{A}\right) ^{2}\psi -2M\boldsymbol{\mu}
  \cdot \mathbf{B}\psi =k^2\psi ,
  \label{dirac17}
\end{equation}
with $k^{2}= 2ME$, $\boldsymbol{\mu}=ge\mathbf{S}/2M$ being the
magnetic moment, where $\mathbf{S}=\boldsymbol{\sigma}/2$ is the spin
operator, and $\psi =\left( \psi _{1},\psi _{2}\right) ^{T}$ is a
two-component spinor.
To describe the AB problem, we specify the magnetic flux tube by
\begin{align}
  e\mathbf{A} = {}
  & - \frac{\alpha }{r}\hat{\boldsymbol{\phi }},  \label{eq:Apoint}\\
  eB = {}
  & -\alpha \frac{\delta (r)}{r} \label{eq:Bpoint},
\end{align}
where $\alpha =\Phi /\Phi_{0}$ is related to the AB flux, with $\Phi$
representing the magnetic flux and $\Phi_{0}=2\pi /e$ the quantum of
magnetic flux.
Thus, considering solutions of the form
\begin{equation}
  \psi _{1}(r,\phi)=\sum_{m=-\infty }^{\infty }\chi_{m}(r)e^{im\phi },
\end{equation}
the radial equation, for each angular momentum sector $m$, is written
as
\begin{equation}
\label{eq:agaxi}
  h \chi_{m}(r) = k^2 \chi_{m}(r),
\end{equation}
where
\begin{equation}
 \label{Hamiltoniano}
  h= h_0+\lambda \frac{\delta \left(r\right) }{r},
\end{equation}
and
\begin{equation}
  h_0=
    -\frac{d^{2}}{dr^{2}}-\frac{1}{r}\frac{d}{dr}
    +\frac{\left( m+\alpha \right) ^{2}}{r^{2}},
\end{equation}
with
\begin{equation}
  \label{eq:lambda}
  \lambda = \frac{\alpha s g}{2},
\end{equation}
where $s=\pm 1$ labels the spin projection and $g$  is the $g$ factor.
The general solution of \eqref{eq:agaxi} for $r\neq 0$ is
\begin{equation}
\chi_{m}(r)=a_{m}J_{| m+\alpha| }(kr)+b_{m}Y_{|
m+\alpha| }(kr), \label{bessel}
\end{equation}
where $J_{\eta}(z)$ and $Y_{\eta}(z)$ are the Bessel functions of the
first and second kinds of fractional order $\eta$, respectively, and
$a_{m}$ and $b_{m}$ are arbitrary constants.

\section{Self-adjoint extension}
\label{sec:sae}

Since the operator presented in the form (\ref{Hamiltoniano}) is
neither continuous nor bounded, it is necessary to adapt the
free-particle Hamiltonian to a well-defined operator capable of
satisfying the self-adjointness conditions.
According to the criteria established in \cite{JMP.1985.26.2520}, we
interpret the operator $h$ as a self-adjoint extension of the free
Hamiltonian $h_0$.
To proceed with the self-adjoint extensions of $h_0$, we decompose the
Hilbert space$\mathcal{H}=L^{2}(\mathbb{R}^{2})$ as
$\mathcal{H}=\mathcal{H}_{r}\otimes\mathcal{H}_{\phi}$, where
$\mathcal{H}_{r}=L^{2}(\mathbb{R}^{+},rdr)$ and
$\mathcal{H}_{\phi}=L^{2}(\mathcal{S}^{1},d\phi)$, with
$\mathcal{S}^{1}$ denoting the unit circle in $\mathbb{R}^{2}$.
The operator $-{\partial^{2}_{\phi}}$ is essentially self-adjoint in
$L^{2}(\mathcal{S}^{1},d\phi)$ \cite{Book.1975.Reed.II}, and because
of this, it is possible to obtain the operator $h_{0}$ in each angular
momentum sector.

Now, using the unitary operator
$V:L^{2}(\mathbb{R}^{+},rdr)\rightarrow L^{2}(\mathbb{R}^{+},dr)$,
given by $(V\xi)(r)=r^{1/2}\xi(r)$, the operator $h_{0}$ can be
transformed into the form
\begin{equation}
  \tilde{h}_{0} = V h_0 V^{-1} =
  -\frac{d^2}{dr^2}
  +
  \left[
    (m+\alpha)^2-\frac{1}{4}
  \right]
  \frac{1}{r^2}
\end{equation}
and all the self-adjoint extensions $h_{0,\nu}$ must satisfy the
boundary condition \cite{Book.2004.Albeverio,JMP.1985.26.2520}
\begin{equation}
  \label{bc}
  \chi_{1}=\nu \chi_{0},
\end{equation}
with
\begin{align}
  \chi_{0}= {}
  &
    \lim_{r \rightarrow 0^{+}} r^{|m+\alpha|}\chi_m (r),\\
  \chi_{1}= {}
  &
    \lim_{r \rightarrow 0^{+}} \frac{1}{r^{|m+\alpha|}}
    \left[\chi_m (r)- \chi_{0}\frac{1}{r^{|m+\alpha|}}\right],
\end{align}
where $-\infty < \nu \le \infty $, and
\begin{equation}
  \label{eq:mvalues}
  |m+\alpha|<1.
\end{equation}
In Eq. \eqref{bc}, $\nu$ is the self-adjoint extension parameter, and
$\nu^{-1/2|m+\alpha|}$ sets the length scale of the point interaction
described by $h_{0,\nu}$ \cite{Book.2004.Albeverio}.
The Friedrichs extension, $\nu=\infty$, represents the free
Hamiltonian with regular wave functions at the origin, $\chi_m(0)=0$.
On the other hand, any finite $|\nu|<\infty$ describes a point
interaction at the flux tube, precisely the situation in which the
boundary condition allows for an $r^{-|m+\alpha|}$ singularity in the
wave functions at the origin.
From the operator-theoretic point of view, the radial Hamiltonian in
the sector $|m+\alpha|<1$ has deficiency indices $(1,1)$ and therefore
admits a one-parameter family of self-adjoint extensions.

\section{Aharonov-Bohm scattering results}
\label{sec:ss}

To solve the scattering problem, we use the general solution
\eqref{bessel}.
Substituting $\chi_{m}(r)$ into the boundary condition \eqref{bc}, we
arrive at the relation
\begin{equation}
  b_{m}=-\tau_{m}^{\nu}(k,\alpha )a_{m}, \label{dmi}
\end{equation}
where
\begin{equation}
  \tau_{m}^{\nu}(k,\alpha )=
  \frac{
     k^{2|m+\alpha|}\Gamma_{-}\sin \left( \pi | m+\alpha|\right)
  }
  {
     k^{2|m+\alpha| }\Gamma_{-}\cos \left( \pi |m+\alpha|\right)
    +4^{|m+\alpha|} \nu \Gamma_{+}}, \label{mul}
\end{equation}
with $\Gamma_{\pm}=\Gamma(1\pm |m+\alpha|)$ and $\Gamma(z)$ the Gamma
function.
Thus, for non-integer flux and any finite $\nu$, we have $b_{m}\neq 0$.
This means that the irregular solutions  $Y_{| m+\alpha|}(kr)$
contributes at the origin whenever the extension differs from the
Friedrichs one, $\nu=\infty$, which is the only member of the family of
with regular wave functions there.
Every $h_{0, \nu}$ is self-adjoint and the irregular component is the
contribution associated with the choice of the boundary condition.

Since the $\delta$-interaction is a short-range interaction and does
not modify the asymptotics beyond the AB phase, the behavior of
$\chi_{m}\left( r\right) $ for $r\rightarrow \infty $ is given by
\cite{JPA.2010.43.354011}
\begin{equation}
   \label{f1asim}
   \chi_{m}(r) \sim \sqrt{\frac{2}{\pi kr}}
   \cos
   \left[
     kr-\frac{|m| \pi}{2}-\frac{\pi}{4}
     +\delta_{m}^{\nu}(k,\alpha)
   \right],
\end{equation}
where $\delta_{m}^{\nu}\left( k,\alpha \right) $ is a scattering phase
shift.
The phase shift is a measure of the difference between the asymptotic
behavior of the full solution and that of the free regular solution
$J_{|m|}(kr)$.
Using the asymptotic behavior of $J_{|m+\alpha|}(kr)$ and
$Y_{|m+\alpha|}(kr)$ given in \cite{Book.1972.Abramowitz} in
(\ref{bessel}), we obtain
\begin{align}
  \label{ass}
  \chi_{m}\left( r\right) \sim {}
  &
    a_{m}\sqrt{\frac{2}{\pi kr}}
    \left[
    \cos \left(
    kr-\frac{\pi}{2}|m+\alpha| -\frac{\pi}{4}\right)
    \right.\nonumber \\
  &
    \left.
    -\tau_{m}^{\nu}(k,\alpha)
    \sin \left(
    kr-\frac{\pi}{2}|m+\alpha| -\frac{\pi}{4}\right) \right].
\end{align}
The direct comparison between \eqref{f1asim} and \eqref{ass} allows us
to identify the quantities
\begin{equation}
  \cos \theta_m^{\nu}=a_{m},
  \qquad
  \sin \theta_m^{\nu}=a_{m}\,\tau_{m}^{\nu}(k,\alpha).
\end{equation}
Therefore, \eqref{f1asim} is satisfied if
\begin{equation}
  a_{m}=
  \left\{
    1+\left[ \tau_{m}^{\nu}(k,\alpha) \right]^{2}
  \right\} ^{-1/2}.
\end{equation}
Now, by comparing the arguments of the cosines above, the following
phase shift is achieved:
\begin{equation}
  \label{eq:phaseshift}
  \delta_{m}^{\nu}(k,\alpha) =
  \delta_{m}^{AB}(\alpha) + \theta_{m}^{\nu}(k,\alpha),
\end{equation}
where
\begin{equation}
\delta_{m}^{AB}(\alpha) =\frac{\pi }{2}(|m|-|m+\alpha|) ,
\end{equation}
is the usual phase shift of the AB scattering, and
\begin{equation}
  \theta_{m}^{\nu}(k,\alpha)=
  \arctan {\left[ \tau_{m}^{\nu}(k,\alpha)\right] }.
\end{equation}
Therefore, the $S$-matrix, $S_{m}^{\nu}(k,\alpha)$, for the
self-adjoint extension is given by
\begin{align}
  \label{eq:Smatrix}
  S_{m}^{\nu}(k,\alpha)
  = {}
  &
    e^{2i\delta_{m}^{\nu}(k,\alpha)}\nonumber \\
  = {}
  &
  e^{2i\delta_{m}^{AB}(\alpha)}
  \left[
    \frac
    {1+i\tau_{m}^{\nu}\left(k,\alpha \right)}
    {1-i\tau_{m}^{\nu}\left(k,\alpha \right)}
  \right].
\end{align}
From the partial-wave analysis, the scattering amplitude can be
written in the form \cite{ruijsenaars1983aharonov,IJMPA.1991.6.3119}
\begin{equation}
  \label{scattering}
  f_\alpha(k,\phi)=
  \sqrt{\frac{1}{2i \pi k}}
  \sum_{m=-\infty}^{+\infty}
  \left[S_{m}^{\nu}(k,\alpha) - 1 \right] e^{im (\phi - \pi)},
\end{equation}
and considering the appropriate
values of $m$ given in \eqref{eq:mvalues}, the amplitude may be
decomposed into regular and irregular contributions, i.e.,
\begin{align}
  \label{eq:fphigen}
  f_\alpha(k,\phi) = {}
  &
    \frac{1}{\sqrt{2i \pi k}}\bigg\{
    \sum_{m\in \mathcal{R}}
    \left[
    e^{2i \delta_{m}^{AB}(\alpha)} - 1
    \right] e^{im (\phi - \pi)}
    \nonumber \\
  &
    +
    \sum_{m\in \mathcal{I}}
    \left[
    e^{2i \delta_{m}^{AB}(\alpha)}
    e^{2i \theta_{m}^{\nu}(k,\alpha)}
    - 1
    \right]
   e^{im (\phi- \pi)}\bigg\},
\end{align}
where $\mathcal{R}$ ($\mathcal{I}$) denotes the summation over $m$ for
$|m+\alpha| \geq 1$ ($|m+\alpha|<1$).
The first sum in Eq. \eqref{eq:fphigen} collects the regular
contributions of the nonsingular channels, while the second arises
from the irregular solution induced by the singular term
$\boldsymbol{\mu} \cdot \mathbf{B}$.
Together, they reproduce the standard AB amplitude plus the correction
generated by the point interaction.

It is instructive to make explicit how our general result relates to
that obtained by Hagen in Ref. \cite{PRL.1990.64.503}.
Thus, we proceed with the decomposition of the magnetic flux as \cite{IJMPA.1991.6.3119,PRD.1993.48.5935}
\begin{equation}
  \alpha = N+\beta,
\end{equation}
with $N $ integer and $0 \leq \beta <1 $.
With this, it can be verified that for each spin state, there is at
most one value of $m$ for which the contribution of the irregular
solution must be taken into account, modifying the scattering
amplitude.
Such values of $m$ are known to be
\cite{IJMPA.1991.6.3119,PRD.1993.48.5935}
\begin{align}
  \label{eq:emes}
  m&=-N,  &  & N   \geq  0, & & s=-1, \nonumber \\
  m&=-N-1,&  & N   \leq -1, & & s=+1.
\end{align}
It is worth noting that the two cases above are precisely those for
which $\alpha s<0$.
Thus, the selection of the singular channel and the attractive
character of the point interaction are one and the same condition.

Hagen's result is retrieved by selecting the particular value of
$\nu = 0$, yielding  $\tau^0_m=\tan(\pi|m+\alpha|)$, and leading to a
sign reversal in the standard AB phase shift
\begin{equation}
  \delta_{m}^{0}(k,\alpha)= \frac{\pi}{2}
  (|m|+|m+\alpha|).
\end{equation}
This shows that the scattering amplitude can be expressed in the form
\begin{align}
  \label{fphi}
  f_{\alpha}(k,\phi)= {}
  &
    f^{AB}(\alpha)+\sqrt{\frac{2i}{\pi k}}
    \sin{(\pi \alpha)} e^{-iN(\phi-\pi)}
    \nonumber \\
  &
    \times e^{-i\phi/2} e^{-i\epsilon(s)\phi/2} \Theta(-\alpha s),
\end{align}
where
\begin{equation}
  f^{AB}(\alpha)=
  -\sqrt{\frac{i}{2\pi k}}\frac{\sin{(\pi \alpha)}
    e^{-iN(\phi-\pi)}}{\cos(\phi/2)} e^{-i\phi/2},
\end{equation}
is the standard AB amplitude \cite{PRD.1993.48.5935}, and we have made
use of the Heaviside step and sign functions $\Theta(x)\equiv
[1+\epsilon(x)]/2$ and $\epsilon(x)\equiv x/|x|$.
The result in \eqref{fphi} is identical to Eq. (13) of
Ref. \cite{PRL.1990.64.503}. Therefore, it is clear that the
scattering amplitude \eqref{eq:fphigen} represents a generalization of
Hagen's result.

\section{Aharonov-Bohm bound state results}
\label{sec:bs}
Having established the general scattering dynamics and noting that
Hagen's result represents a particular case, we now determine the
conditions under which a bound state can emerge.
Our goal is to identify the critical boundary separating the
scattering and bound state regimes in the spin-$1/2$ AB problem.
Bound states can be analyzed by the analytic continuation of the
$S$-matrix to the complex plane, and it is well-known that its poles
in the positive imaginary $k$-axis correspond to bound states
\cite{Book.2011.Sakurai}.
Here, this condition is met when the denominator of \eqref{eq:Smatrix}
vanishes with the substitution $k \to i \kappa$, with $\kappa^2 = - 2
M E_b$ and $E_b <0$.
Thus, the bound state energy is uniquely determined by
\begin{equation}
  \label{eq:bsenergybg}
  E_b =
  - \frac{2}{M}
  \left[
    - \nu  \frac{\Gamma_{+}}{\Gamma_{-}}
    % \frac{\Gamma(1+|m+\alpha|)}{\Gamma(1-|m+\alpha|)}
  \right]^{{1}/{|m+\alpha|}}.
\end{equation}
Since the radial operator has deficiency indices $(1,1)$ in the sector
$|m+\alpha|<1$, the self-adjoint extension has at most one negative
eigenvalue, and Eq. \eqref{eq:bsenergybg} accounts for the single pole
of Eq. \eqref{eq:Smatrix} on the positive imaginary $k$ axis. The
singular channel therefore supports at most one bound state.

\subsection{Physical determination of the extension parameter}

In the preceding analysis of scattering, the self-adjoint extension
parameter $\nu$ was treated as a free parameter characterizing the
interaction at the origin.
In a rigorous mathematical sense, different physics inside the flux tube
will be described by different Hamiltonians, given as a certain
self-adjoint extension of a formal differential operator.
However, for the model to describe a physical system, $\nu$ must be
determined by the fundamental properties of the interaction itself.
Here, we adopt the regularization procedure due to Kay and
Studer (KS) \cite{CMP.1991.139.103}, in which the point interaction is
modeled by a solenoid of finite radius $R$, and the self-adjoint
extension parameter is determined by matching the logarithmic
derivative of the static solution (i.e., zero energy solution) of the
\emph{regularized} problem to that of the solution parametrized by the
extension.

We therefore replace the idealized flux line described by Eqs.  \eqref{eq:Apoint}--\eqref{eq:Bpoint} with a solenoid of finite 
radius $R$,
\begin{align}
  e\mathbf{A}_{R} = {}
  & -\frac{\alpha}{r}\,\Theta(r-R)\,\hat{\boldsymbol{\phi}},\\
  eB_{R} = {}
  & -\frac{\alpha}{R}\,\delta(r-R),
  \label{eq:regularization}
\end{align}
in such a way that we replace $\delta(r)/r$ by $\delta(r-R)/R$ in
Eq. \eqref{Hamiltoniano}.
The regularized tube carries the same total flux and reduces to Eqs.
\eqref{eq:Apoint} and \eqref{eq:Bpoint} as $R \to 0^{+}$.
The essential feature of the above equations is that the interior of
the tube is \emph{flux free}.
Let $\tilde{\chi}_{\rm in}(r)$ be the static solution for $h$ in
\eqref{eq:agaxi} with the regularization in \eqref{eq:regularization},
and $\tilde\chi_{\rm out}(r)$ be the bound state solution for $h_0$.
One therefore has,
\begin{equation}
  \left[
    -\frac{d^{2}}{dr^{2}}-\frac{1}{r}\frac{d}{dr}
    +\frac{m^{2}}{r^{2}}
    +\frac{\lambda}{R}\delta(r-R)
  \right]\tilde\chi_{\rm in}(r)=0, \quad r<R,
  \label{eq:incore}
\end{equation}
and
\begin{equation}
  \left[
     \frac{d^{2}}{dr^{2}}+\frac{1}{r}\frac{d}{dr}
     -\left(\frac{|m+\alpha|^{2}}{r^{2}}+\kappa^2\right)
     \right]\tilde\chi_{\rm out}(r)=0, \quad r>R.
  \label{eq:outcore}
\end{equation}
The bound state solution selected by the physics inside the core,
which, as will be shown below, is associated with the self-adjoint
extension parameter $\nu$ and is determined by
($\tilde\chi'\equiv d\tilde\chi/dr$) \cite{CMP.1991.139.103}
\begin{equation}
  \label{eq:ksbc}
  \lim_{R \to 0^+} R
  \left.\frac{\tilde\chi_{\rm in}'(r)}{\tilde\chi_{\rm in}(r)}\right|_{r=R}=
  \lim_{R \to 0^+} R
  \left.\frac{\tilde\chi_{\rm out}'(r)}{\tilde\chi_{\rm out}(r)}\right|_{r=R}.
\end{equation}
The left-hand side of \eqref{eq:ksbc} can be obtained by multiplying
Eq. \eqref{eq:incore}  by $r$ and integrating from $0$ to $R^{+}$, ensuring
that the shell is entirely closed,
\begin{equation}
  \label{eq:intincore}
  \int_{0}^{R^{+}} \frac{d}{dr}\left[r\tilde\chi_{\rm in}'(r)\right] dr=
  m^2\int_0^{R^+}\frac{\tilde\chi_{\rm in}(r)}{r} dr
  +\lambda\tilde\chi_{\rm in}(R).
\end{equation}
The solution of \eqref{eq:incore} is regular at the origin and behaves
as $\tilde\chi_{\rm in}(r) = C r^{|m|}$, so the boundary term in the
first integral vanishes, and for $|m|\neq 0$, the second integral is
given by
\begin{equation}
  \int_0^R\frac{\tilde\chi_{\rm in}(r)}{r}dr
  = C \int_0^R r^{|m|-1}dr
  = \frac{\tilde\chi_{\rm in}(R)}{|m|}.
  \end{equation}
Therefore, dividing \eqref{eq:intincore} by $\tilde\chi_{\rm in}(R)$,
one finds
\begin{equation}
  \label{eq:lhsks}
  \lim_{R\to 0^+}
  R \left.\frac{\tilde\chi_{\rm in}'(r)}{\tilde\chi_{\rm in}(r)}\right|_{r=R}
  =|m|+\lambda.
\end{equation}
For $|m|=0$, the centrifugal term is absent from Eq. \eqref{eq:incore},
and the result above still holds.
The result in \eqref{eq:lhsks} differs from those in
Refs. \cite{PRD.2012.85.041701,AoP.339.510.2013}, in which the second
integral in \eqref{eq:intincore} was argued to vanish when $R \to 0$.
The integral indeed tends to zero, but it should be compared to
$\tilde\chi_{\rm in}(R)$ itself in this limit.
The result obtained here coincides with the results in
\cite{PRD.2012.85.041701,AoP.339.510.2013} only for the fundamental
inner core sector $|m|=0$, that is, when the interior solution is
taken to be flat.

It is interesting to observe that the same result is obtained from the
observation that the logarithmic derivative provides $R
\tilde\chi'/\chi=|m|$ for $r<R$, and the regularized $\delta$ adds
$\lambda$ through the discontinuity or jump condition
$\tilde\chi'(R^{+})-\tilde\chi'(R^{-})=(\lambda/R)\tilde\chi(R)$, which
is obtained from the standard procedure of integrating
Eq. \eqref{eq:incore} in the interval $[R-\epsilon,R+\epsilon]$ and
taking the limit $\epsilon \to 0$ at the end.
This latter approach is exactly what was adopted by Hagen in
Ref. \cite{PRL.1990.64.503}.

The right-hand side of \eqref{eq:ksbc} is obtained from the solution
of Eq. \eqref{eq:outcore}.
The general solution is given in terms of the modified Bessel functions:
\begin{equation}
  \tilde\chi_{\rm out}(r)=
  A  K_{|m+\alpha|}(\kappa r) + B I_{|m+\alpha|}(\kappa r),
\end{equation}
with $A$ and $B$ constants.
The normalization condition imposes $B=0$ as $I_{|m+\alpha|}(\kappa r)$
diverges at infinity.
So, we have
\begin{equation}
  R \frac{\tilde\chi_{\rm out}'(r)}{\tilde\chi_{\rm out}(r)}=
  (\kappa R) \frac{K_{|m+\alpha|}'(\kappa r)}{K_{|m+\alpha|}(\kappa r)},
\end{equation}
which, upon using the asymptotic form for the Bessel functions, becomes
\begin{equation}
  \label{eq:rhsks}
  \lim_{R \to 0^{+}}
  \left.
  R\frac{\tilde\chi_{\rm out}'(r)}{\tilde\chi_{\rm out}(r)}
  \right|_{r=R}
  =|m+\alpha|\frac{\Omega-1}{\Omega+1},
\end{equation}
with
\begin{equation}
  \Omega = -\left(\frac{\kappa R}{2}\right)^{2|m+\alpha|}
  \frac{\Gamma_{-}}{\Gamma_{+}}.
\end{equation}
Equating Eqs. \eqref{eq:lhsks} and \eqref{eq:rhsks}, one finds
\begin{equation}
  \label{eq:bsenergyks}
  E_b=-\frac{2}{MR^2}
  \left[
    \frac{|m|+\lambda+|m+\alpha|}{|m|+\lambda-|m+\alpha|}
    \frac{\Gamma_{+}}{\Gamma_{-}}
  \right]^{1/|m+\alpha|},
\end{equation}
which gives the bound-state energy from the KS approach.
Comparing the energies in Eqs. \eqref{eq:bsenergybg} and
\eqref{eq:bsenergyks} and solving for $\nu$, we obtain a direct
connection between the self-adjoint extension parameter and the
physical parameters
\begin{equation}
  \nu = -\frac{1}{R^{2|m+\alpha|}}
  \left(
    \frac{|m|+\lambda + |m+\alpha|}{|m|+\lambda - |m+\alpha|}
  \right).
  \label{eq:ksnu}
\end{equation}

\begin{figure}[t]
  \centering
  \includegraphics[width=\columnwidth]{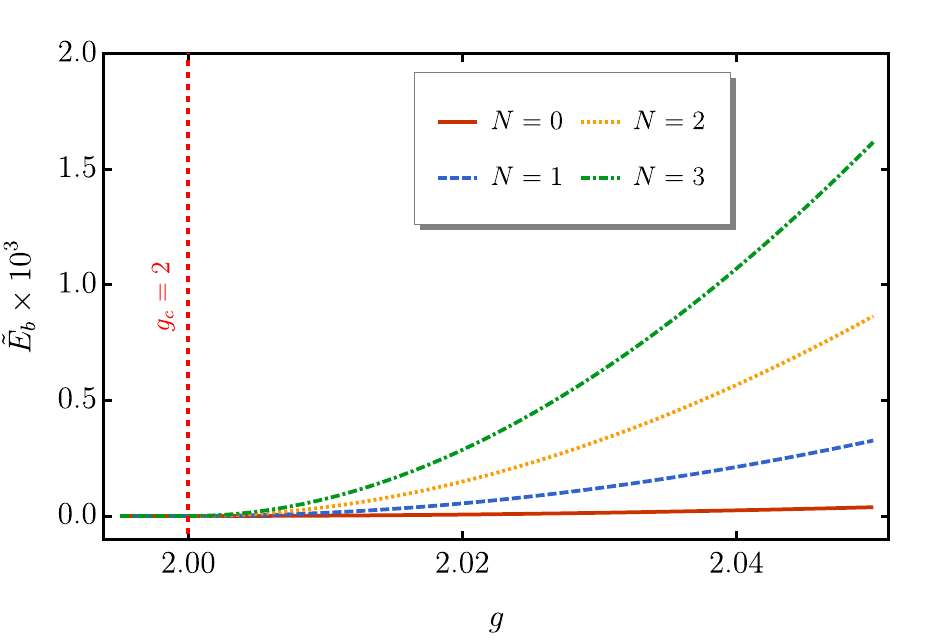}
  \caption{
    Scaled binding energy $\tilde{E}_b$ of the singular channel as a
    function of $g$, for $\beta=1/2$ and several flux sectors.
    All sectors share the threshold $g_c=2$, but the binding deepens
    with $|N|$.}
  \label{fig:fig1}
\end{figure}

\subsection{Binding threshold and bound-state energy}

The existence of bound states, which are solutions with negative
energy ($E_b<0$), requires an effectively attractive interaction that
implies $\lambda<0$. As noted after Eq. \eqref{eq:emes}, this is
automatically satisfied in the singular channels, where $\alpha s<0$.
Writing $\lambda=-|\alpha|g/2$ and introducing
\begin{equation}
\label{eq:tbetas}
 \tilde{\beta} =
  \begin{cases}
    \beta, &\text{ for the channel } m=-N, \;\;\;\;\;\;\; N\geq 0. \\
    1-\beta, & \text{ for the channel } m = -N -1, \; N \leq -1,
  \end{cases}
\end{equation}
so that $|m+\alpha|=\tilde\beta$, one verifies directly from
Eq. \eqref{eq:emes} the identity
$|m|=|\alpha|-\tilde\beta$, which implies
\begin{equation}
  \label{eq:identity}
  |m|+|m+\alpha| = |\alpha|.
\end{equation}
That is, in the singular channels, the interior centrifugal index and
the exterior AB index add up exactly to the flux.
Thus, using $\lambda+|m|+|m+\alpha| = -|\alpha|(g-2)/2$,
Eq. \eqref{eq:ksnu} therefore becomes
\begin{equation}
  \label{eq:nug}
  \nu = -\frac{1}{R^{2\tilde{\beta}}}\,
  \frac{g-2}{(g-2)+4\tilde{\beta}/|N+\beta|} .
\end{equation}
A bound state requires $\nu<0$, so that the bracket in
Eq. \eqref{eq:bsenergybg} is positive.
For $g>2$ the numerator and the denominator of \eqref{eq:nug} are both
positive, and the critical magnetic $g$ factor is
\begin{equation}
  \label{eq:gc}
  g_c = 2.
\end{equation}
for every flux sector $\alpha = N+\beta$.
This is a key result of the present work.
The threshold for binding depends neither on the integer part $N$ of
the flux, nor on its fractional part $\beta$, nor on the
regularization radius $R$, and $\nu=0$, the extension that reproduces
Hagen's solution, occurs precisely at $g=2$ in every channel.
Since the Zeeman coupling $\lambda=-|\alpha|g/2$ grows linearly with
the flux, one might expect a large flux to compensate for a small 
$g$ and produce binding below $g=2$.
It does not, because the centrifugal barrier
$|m|=|\alpha|-\tilde\beta$ grows at exactly the same rate, as
expressed by the identity \eqref{eq:identity}.
Attraction and repulsion scale together, and their competition is
decided by $g$ alone.
Figure \ref{fig:fig1} shows the resulting scaled binding energy as a
function of $g$ for several flux sectors: all of them share the
threshold $g_c=2$, but the state becomes more deeply bound as $|N|$
grows.

It is instructive to examine in more detail what the choice $\nu=0$,
which reproduces Hagen's result in \cite{PRL.1990.64.503}, means
within the regularization adopted here.
By Eq. \eqref{eq:nug}, it corresponds to $g=2$ for arbitrary flux.
Far from being an arbitrary member of the family of admissible
extensions, $\nu=0$ is therefore the member selected by the physical
regularization whenever the magnetic moment takes its Dirac value.
Hagen's conclusion that a Dirac particle is unbound is thus confirmed
and, at the same time, sharpened: it is not a statement about one
particular boundary condition among many, but about the boundary
condition that the microscopic structure of the flux tube actually
imposes.
An anomalous magnetic moment ($g>g_c$) is a necessary condition for
the appearance of a bound state in the singular channels.

\begin{figure}[t]
  \centering
  \includegraphics[width=\columnwidth]{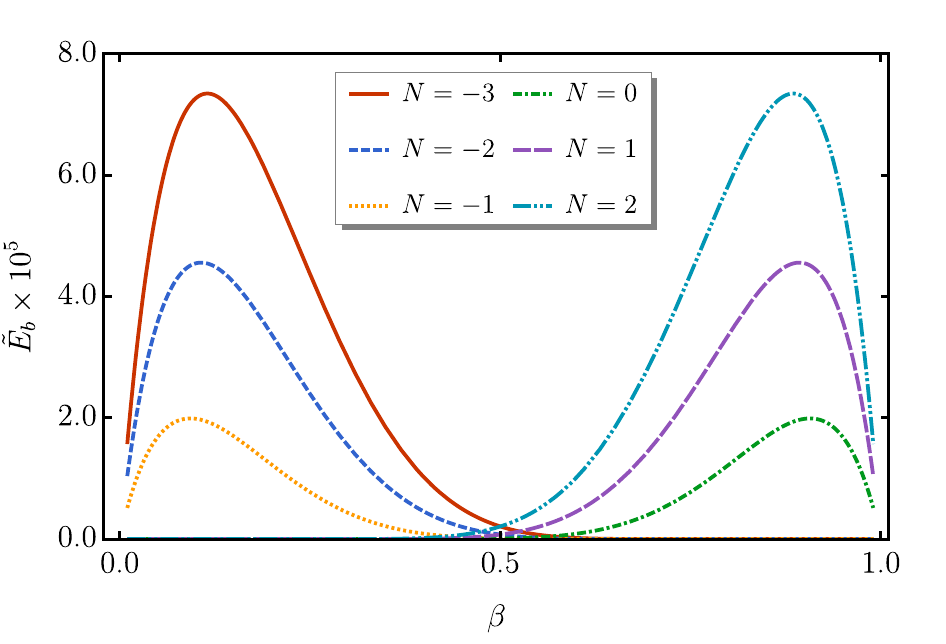}
  \caption{
    Scaled binding energy $\tilde{E}_b$ as a function of the fractional
    part $\beta$ of the magnetic flux for different values of $N$ with
    $g=g_e$.
    We can observe the symmetry between the two families for $N \ge 0$
    and $N<0$, which are related by $\beta \to 1-\beta$, $N\to-N-1$.}
  \label{fig:fig2}
\end{figure}

Factoring out the regularization scale $2/(MR^2)$ and substituting
Eq. \eqref{eq:nug} into Eq. \eqref{eq:bsenergyks} yields the
dimensionless scaled binding energy $\tilde{E}_b=-MR^2E_b/2$,
\begin{equation}
  \label{eq:bsenergy_gc}
  \tilde{E}_b =
  \left[
    \frac{g-2}{(g-2)+4\tilde{\beta}/|N+\beta|}
    \frac{\Gamma(1+\tilde{\beta})}{\Gamma(1-\tilde{\beta})}
  \right]^{{1}/{\tilde{\beta}}}.
\end{equation}
In the fundamental sectors, where $|N+\beta|=\tilde\beta$, the bracket
reduces to $(g-2)/(g+2)$.
For $|N| \ge 1$, the quantity $4\tilde\beta/|N+\beta|$ decreases as the
flux grows, so the bracket increases and the state becomes more deeply
bound: the flux sector does not move the threshold, but it does
control how fast the binding energy grows once the threshold is
crossed.
We note that the fractional part of the AB flux determines the
behavior of the bound state energy, and there is an interesting
symmetry between the singular channels with $\beta \to 1-\beta$, which
pairs the sector $N$ with $-N-1$.
Since $\tilde{E}_b$ carries no explicit dependence on $M$ or $R$, it
depends on the physical parameters of the problem only through $g$ and
$\tilde{\beta}$, as well as on the flux sector. 
Therefore, it is insensitive to the microscopic regularization scale.
In contrast, the physical binding energy $E_b$ retains an
inverse-square dependence on the short-distance scale.
Figure \ref{fig:fig2} illustrates the behavior of the scaled binding
energy $\tilde{E}_b$ for the physical electron with an anomalous
magnetic $g$ factor $g=g_e\cong 2.002 32$ \cite{RMP.2.025002.2025} as
a function of the fractional part $\beta$ for different values of $N$.
%Exact value of g_e=2.002 319 304 360 92(36)

\section{Experimental implications and outlook}
\label{sec:exp}
Our analysis has direct implications for mesoscopic AB interferometers
based on semiconductor or metallic rings, where phase-coherent
transport is routinely probed, and spin-polarized carriers are
naturally present.
Early experiments on normal-metal rings established the robustness of
$h/e$ AB oscillations in magnetoresistance at low temperatures
\cite{PhysRevLett.54.2696,PhysRevLett.55.1610}, while more recent work
on semiconductor quantum rings with Rashba spin-orbit coupling has
demonstrated the controlled manipulation of spin-dependent geometric
phases \cite{Nagasawa2013NatCommun}.
In such devices, however, the observed phenomena are primarily
interference and geometric-phase effects rather than a direct
isolation of the singular short-distance regime considered in the
present work.
In our framework, the flux tube is modeled by a localized core, and
the spin experiences a Zeeman coupling whose strength depends on $g$
and local fields.
Within the finite radius regularization considered here, whenever the
universal condition $g > 2$ is satisfied, the relevant singular
channel supports a bound state in addition to the usual AB scattering.

In a mesoscopic context, the bound state itself lies below the
scattering continuum and does not, by itself, appear as a feature in
the conductance.
What makes the singular channel observable is the resonance that
accompanies it.
It has been shown that whenever a bound state survives the $R \to 0$
limit of the AB potential, it is always accompanied by a resonance
whose energy scales with the modulus of the binding energy and whose
line shape is not Breit-Wigner \cite{Moroz1996}.
Within our framework, the universal condition $g>2$ is precisely the
criterion controlling the appearance of this pair.
In an interferometer geometry in which the flux core is embedded in
one arm of an AB ring, the second arm provides a nonresonant path that
interferes with the resonant one, so that the feature is expected to
appear as an asymmetric, Fano-like line shape in the conductance,
accompanied by an abrupt shift of the transmission phase.
A quantitative treatment would require specifying how the flux core is
coupled to the transport path, which lies beyond the scope of the
present work.

The most distinctive signature, however, is its spin selectivity.
As shown after Eq. \eqref{eq:emes}, the singular channel exists only
for $\alpha s<0$, so that for a given sign of the flux, the bound
state, and hence the accompanying resonance, occurs in a single spin
projection.
Reversing the flux transfers the feature to the opposite spin channel.
In suitably engineered spin-resolved AB devices or in quantum rings
with strong spin-orbit interaction, this asymmetry between the two
spin channels would provide a direct fingerprint of the singular
Zeeman interaction, superimposed on the standard AB oscillations.

\section{Conclusion}
\label{sec:conc}
In this work, we have revisited the spin-$1/2$ AB problem by combining
the self-adjoint extension formalism with a finite-radius
regularization of the magnetic flux tube.
The regularization uniquely selects the physical self-adjoint
extension and thereby provides a direct physical interpretation for
the boundary condition governing the singular Zeeman interaction.
In this framework, Hagen's scattering solution corresponds to the
extension $\nu=0$, which leads exactly to $g=2$ in every flux sector.
The binding threshold is therefore universal, $g_c=2$, independent of
the flux sector and the regularization radius, and a bound state
exists in the singular channels precisely when $g>2$.
Hagen's conclusion is thus confirmed and, at the same time, sharpened:
it is not a statement about one arbitrary member of a family of
extensions, but about the member that physics actually selects
when the magnetic moment takes its Dirac value.
An anomalous magnetic moment is a necessary condition for the
formation of a bound state in the singular channels.
What the flux sector controls is the depth of the state: the
closed-form binding energy \eqref{eq:bsenergy_gc} shows a marked
enhancement with the integer part of the flux, so that a given anomaly
binds far more strongly at large flux.
Finally, although the AB effect itself is experimentally well
established, the short-distance regime investigated here has not, to
our knowledge, been isolated experimentally.
We therefore expect that the present results may be relevant not only
to the spin-$1/2$ AB problem itself but also to a broader class of
systems with singular gauge fields and topological defects, where the
interplay between singular interactions and microscopic boundary
conditions may become experimentally accessible and play a decisive
role in determining the quantum spectrum.

\section*{Acknowledgments}
%V.S. acknowledges the Spanish MCIN for funding from the European Union Next Generation EU (PRTRC17.I1) and Consejería de Educación from Junta de Castilla y León through the QCAYLE project and grant No. PID2023-148409NS-i00 MTM funded by AEI/10.13039/501100011033, and RED2022-134301-T.
This work was partially supported by the Brazilian agencies CAPES,
CNPq, and FAPEMA.
E.O.S. acknowledges support from grants CNPq/305427/2026-1, FAPEMA/UNIVERSAL-06395/22, and CAPES/Code 001.
F.M.A. acknowledges CNPq Grant No. 313124/2023-0 and Fundação Araucária
Project No. 305.\\
%\end{acknowledgments}

\paragraph*{Conflict of interest.}%
The authors declare no conflicts of interest.\\

\paragraph*{Data availability.}
There are no publicly available research data or software supporting this manuscript. Requests for further information or data should be sent to the authors.

%\bibliographystyle{apsrev4-2}
%\bibliography{bib}

\end{document}